\documentclass[journal]{IEEEtran}
\usepackage{multirow}
\usepackage{xcolor}
\usepackage{amsmath}
\usepackage{booktabs}
\usepackage{cite}
\ifCLASSINFOpdf
   \usepackage[pdftex]{graphicx}
\else
\fi
\begin{document}
%
% paper title
% Titles are generally capitalized except for words such as a, an, and, as,
% at, but, by, for, in, nor, of, on, or, the, to and up, which are usually
% not capitalized unless they are the first or last word of the title.
% Linebreaks \\ can be used within to get better formatting as desired.
% Do not put math or special symbols in the title.
\title{Nanophotonic cavity based synapse for scalable photonic neural networks}

% author names and IEEE memberships
% note positions of commas and nonbreaking spaces ( ~ ) LaTeX will not break
% a structure at a ~ so this keeps an author's name from being broken across
% two lines.
% use \thanks{} to gain access to the first footnote area
% a separate \thanks must be used for each paragraph as LaTeX2e's \thanks
% was not built to handle multiple paragraphs
%

\author{Aashu~Jha,
        Chaoran~Huang,%~\IEEEmembership{Member,~IEEE,}
        ~Thomas Ferreira deLima,
        ~Hsuan-Tung Peng,
        ~Bhavin Shastri,~\IEEEmembership{Senior~Member,~IEEE}
        ~Paul~R.~Prucnal,~\IEEEmembership{Life~Fellow,~IEEE}% <-this % stops a space
\thanks{This work was supported by the Office of Naval Research (Award N00014-18-1-2297). \emph{(Corresponding author: Aashu Jha)}. A. Jha, C. Huang, T.F. deLima, H.T.Peng, P. R. Prucnal are with the Department of Electrical and Computer Engineering, Princeton University, Princeton, NJ 08544, USA. C. Huang is also with the Department of Electronic Engineering at the Chinese University of Hong Kong. B. Shastri is with the Department of Physics, Engineering Physics and Astronomy, Queens University, Canada. }
}

% note the % following the last \IEEEmembership and also \thanks - 
% these prevent an unwanted space from occurring between the last author name
% and the end of the author line. i.e., if you had this:
% 
% \author{....lastname \thanks{...} \thanks{...} }
%                     ^------------^------------^----Do not want these spaces!
%
% a space would be appended to the last name and could cause every name on that
% line to be shifted left slightly. This is one of those "LaTeX things". For
% instance, "\textbf{A} \textbf{B}" will typeset as "A B" not "AB". To get
% "AB" then you have to do: "\textbf{A}\textbf{B}"
% \thanks is no different in this regard, so shield the last } of each \thanks
% that ends a line with a % and do not let a space in before the next \thanks.
% Spaces after \IEEEmembership other than the last one are OK (and needed) as
% you are supposed to have spaces between the names. For what it is worth,
% this is a minor point as most people would not even notice if the said evil
% space somehow managed to creep in.

% The paper headers
\markboth{Journal of \LaTeX\ Class Files,~Vol.~14, No.~8, August~2015}%
{Shell \MakeLowercase{\textit{et al.}}: Bare Demo of IEEEtran.cls for IEEE Journals}
% The only time the second header will appear is for the odd numbered pages
% after the title page when using the twoside option.
% 
% *** Note that you probably will NOT want to include the author's ***
% *** name in the headers of peer review papers.                   ***
% You can use \ifCLASSOPTIONpeerreview for conditional compilation here if
% you desire.

% If you want to put a publisher's ID mark on the page you can do it like
% this:
%\IEEEpubid{0000--0000/00\$00.00~\copyright~2015 IEEE}
% Remember, if you use this you must call \IEEEpubidadjcol in the second
% column for its text to clear the IEEEpubid mark.

% use for special paper notices
%\IEEEspecialpapernotice{(Invited Paper)}

% make the title area
\maketitle

% As a general rule, do not put math, special symbols or citations
% in the abstract or keywords.
\begin{abstract}
The bandwidth and energy demands of neural networks has spurred tremendous interest in developing novel neuromorphic hardware, including photonic integrated circuits. Although an optical waveguide can accommodate hundreds of channels with THz bandwidth, the channel count of photonic systems is always bottlenecked by the devices within. In WDM-based photonic neural networks, the synapses, i.e. network interconnections, are typically realized by microring resonators (MRRs), where the WDM channel count ($N$) is bounded by the free-spectral range of the MRRs. For typical Si MRRs, we estimate $N \leq 30$ within the C-band. This not only restrains the aggregate throughput of the neural network but also makes applications with high input dimensions unfeasible.  We experimentally demonstrate that photonic crystal nanobeam based synapses can be FSR-free within C-band, eliminating the bound on channel count. This increases data throughput as well as enables applications with high-dimensional inputs like natural language processing and high resolution image processing. In addition, the smaller physical footprint of photonic crystal nanobeam cavities offers higher tuning energy efficiency and a higher compute density than MRRs. Nanophotonic cavity based synapse thus offers a path towards realizing highly scalable photonic neural networks. 
% bringing the energy per operation down to 0.02 pJ
\end{abstract}

% no keywords

% For peer review papers, you can put extra information on the cover
% page as needed:
% \ifCLASSOPTIONpeerreview
% \begin{center} \bfseries EDICS Category: 3-BBND \end{center}
% \fi
%
% For peerreview papers, this IEEEtran command inserts a page break and
% creates the second title. It will be ignored for other modes.
\IEEEpeerreviewmaketitle

\section{Introduction}

\begin{figure}
    \centering
    \includegraphics[width = 0.4\textwidth]{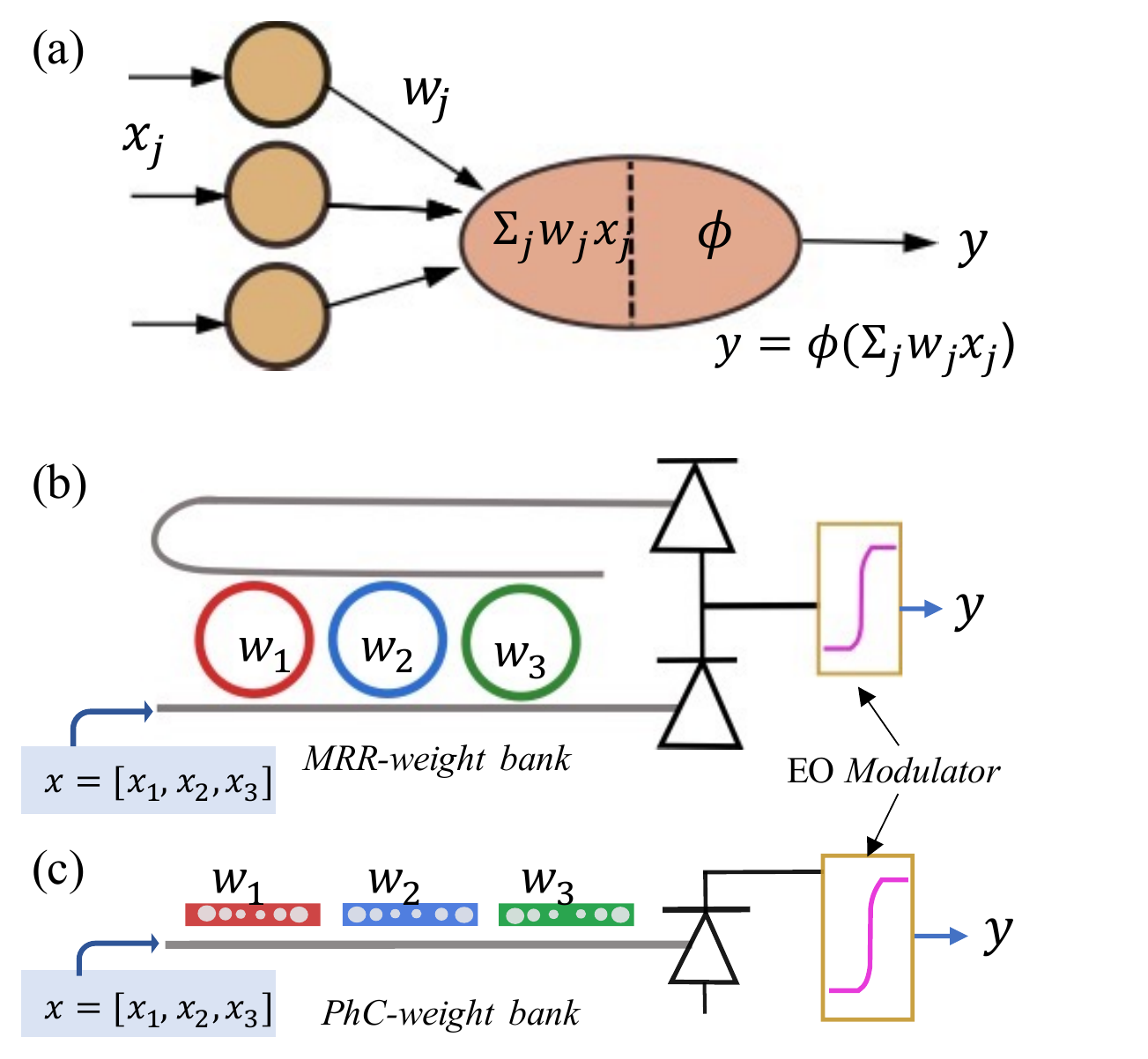}
    \caption{(a) Model of the neuron showing linear weighting and summing of inputs, followed by a nonlinear function, $\phi$, (b) WDM compatible MRR-based neuron with balanced photodetector and an electro-optic (EO) modulator, and (c) a PhC-based neuron with photodetector and an EO modulator. Photodetector carries out the summing operation while the EO modulator relays the nonlinear transformation. }
    \label{fig:model}
\end{figure}

Artificial intelligence (AI) enabled by neural networks has seen an explosive growth in the last decade \cite{amodeiai}. Neural network algorithms are inherently parallel and thus inefficient on conventional hardware where processing is sequential, this challenge has motivated neuromorphic hardware engineering \cite{shastri2021photonics}. In neuromorphic electronics, bandwidth and interconnectivity have to be traded off \cite{shastri2021photonics}. Photonics offers the opportunity to simultaneously achieve high bandwidth with high interconnectivity, in tandem with low power consumption and low latency \cite{prucnal2017neuromorphic}. The interconnections between ``neurons" in a neural network are known as synapses.  Figure \ref{fig:model}(a) shows the functional model of a neuron: at each neuron, inputs from other neurons are \emph{linearly} weighed by synapses, summed and nonlinearly transformed, generating an output. In photonics, one can theoretically parallelize hundreds of high speed information channels onto a single waveguide by encoding each channel on a different wavelength, a scheme known as wavelength-division multiplexing (WDM) \cite{1073653}. Photonic implementations of synapses include coherent \cite{shen2017deep} and WDM \cite{tait2016microring} systems. WDM maximally utilizes the high bandwidth of photonics, which then translates to significant network benefits in terms of information throughput. Ref. \cite{tait2016microring} introduced the concept of a microring weight bank, shown in Fig. \ref{fig:model}(b), which is a WDM-based synapse where an array of microring resonators (MRRs) selectively weighs an input vector of wavelength channels. Each MRR addresses a given wavelength channel where the transmission corresponding to that wavelength is the weight for the synapse. The optical output from the pass and drop ports of the weight bank are then summed and differentially detected by a pair of balanced photodetectors. The summed electrical signal from the PD then drives an  electro-optical modulator with an optical pump to generate a nonlinearly transformed optical output \cite{tait2019silicon}.

\begin{figure}
\centering
    \includegraphics[width = 0.4\textwidth]{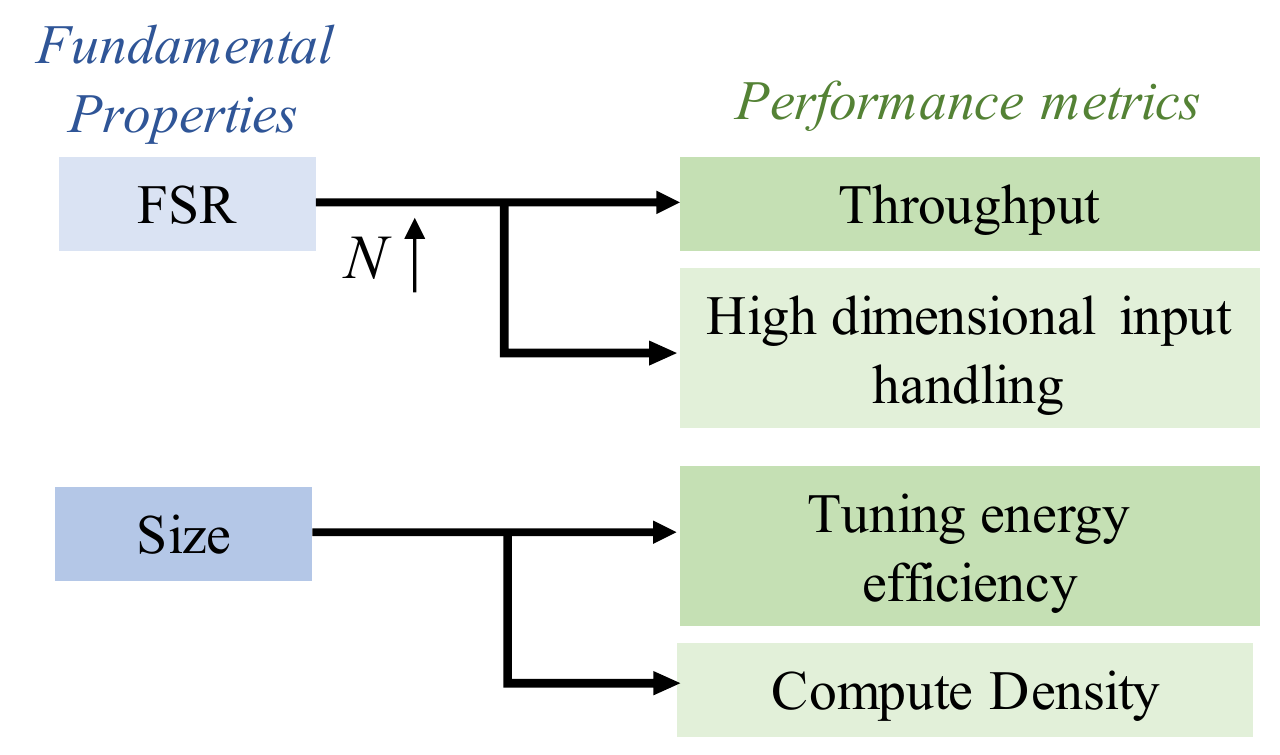}
    \caption{Performance benefits of PhCs as synapses. Their fundamental properties, such as FSR-free operation and small physical footprint, translates to improvements in system performance metrics, including throughput, ability to handle high-dimensional input applications, higher tuning energy efficiency and high compute density. $N$ refers to the WDM channel count.}
    \label{fig:perf}
\end{figure}

Weighting at synapses represents the majority of operations in a neural network, the overall network performance is thus heavily dependent on the synapse properties, as encapsulated in Figure \ref{fig:perf}. The free-spectral range (FSR) of an MRR limits the number of wavelength channels, $N$, i.e. the number of synapses in a network. We estimate $N\leq30$ for silicon MRRs. The constraint on channel count inevitably limits the throughput, i.e. the amount of information being processed, as well constrains the input dimension making some applications unfeasible. Ref. \cite{eid2016fsr} shows an approach to eliminate the FSR constraint by suppressing all but one microring resonance but the level of suppression varies with wavelength which isn't ideal.  In practice, $N$ is even lower due to effects like inter-channel crosstalk, and there have been efforts to mitigate it \cite{tait2018two}. Besides spectral scale, the physical scale of the network also depends on the synapses. A smaller synapse improves the compute density, which is a measure of the number of operations done per unit time normalized per unit area \cite{nahmias2019photonic}. The size of a photonic device used as a synapse determines the energy consumed while tuning the weight -- the smaller the device, the higher the efficiency. While MRRs are relatively compact, as compared to interferometers \cite{shen2017deep}, you can only make it so much smaller without introducing too much bending loss, typically MRRs have radii larger than 5$\mu$m. To make the synapse smaller entails a novel synapse.

Here, we present a novel WDM synapse -- a photonic crystal (PhC) weight bank, as illustrated in Figure \ref{fig:model}(c). Photonic crystals (PhCs) are nanophotonic devices with periodic dielectric function that allows to manipulate the propagation of light. By introducing defects or breaking the periodicity is some way, localized modes for some light frequency can be formed \cite{joannopoulos1995photonic} forming the basis of engineering optical resonators. They are also widely popular for nonlinear photonic applications like all-optical switching \cite{yang2007observation, tanabe2005all}. The simplest form of PhCs is a one-dimensional PhC, also known in literature as nanobeam cavities. Their one-dimensional nature means they can easily be implemented in photonic integrated circuits where light can be evanescently coupled through a bus waveguide. PhC nanobeam based devices have previously been used as wavelength filters \cite{zhou2017compact}. Here, we propose a PhC-based synapse, which replaces the MRRs in the current weight bank architecture, for two reasons. Firstly, we show that PhCs can be FSR-free, which significantly improves the WDM channel count, thereby enhancing the system throughput and enabling neural network applications with high-dimensional inputs. Secondly, we show that the physical size reduction with PhCs improves the overall tuning energy efficiency. Our results show that photonic crystals have the potential to enable the most scalable next-generation photonic neural networks with high throughput, energy efficiency and compute density. 
% Nanophotonic devices such as photonic crystals have long been commended for their nonlinear processing efficiency, owing to their ultra-small mode volume \cite{nozaki2010sub} that enables enhanced light-matter interaction. This feature enables higher resonance tuning efficiency relative to MRRs \cite{frank2010programmable, schmidt2007compact, dong2018ultra}. It is also possible to design photonic crystals with much longer FSR than MRRs which directly improves the permissible channel count and thus expands the system throughput. Photonic crystal nanobeam cavities occupy the same area as an optical waveguide which means significantly higher compute density i.e. processing done per unit area, thereby resulting in a significant reduction of system size over MRR-based systems. With such evidence of the fundamental superiority of photonic crystals, it is natural to implement them as synapses in photonic neural network hardware.

\section{Scaling challenge in MRR-based synapses}

\begin{figure}[h]
    \centering
    \includegraphics[width = 0.5\textwidth]{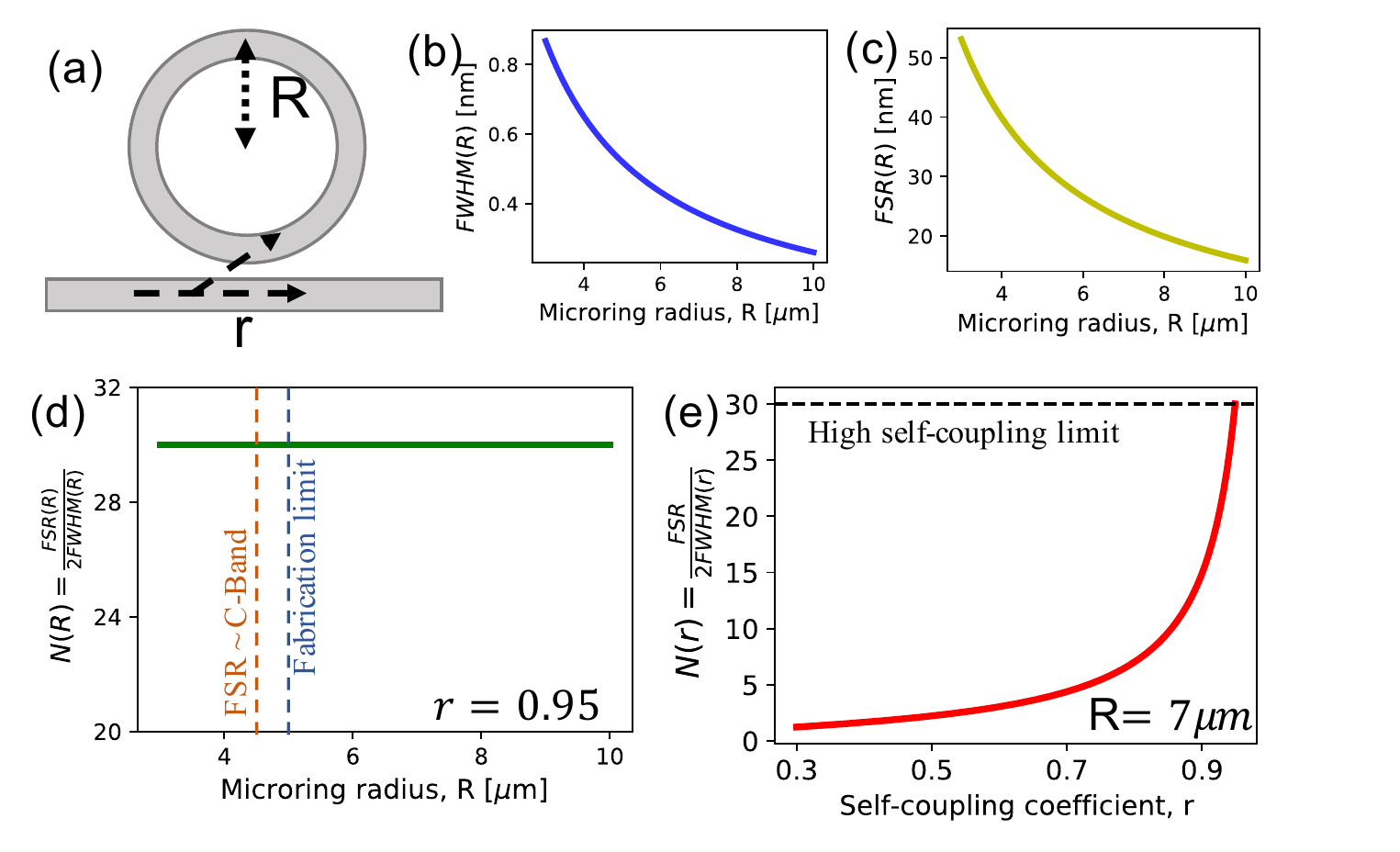}
    \caption{WDM channel count limitation of microring resonator based WDM systems, (a) Illustration of a microring of radius $R$ and coupling coefficient, $r$, (b) full-width half-maximum (FWHM) of MRR resonance as a function of radius $R$, (c) free-spectral range (FSR) of MRR as a function of radius $R$, (d) WDM channel count $N$, calculated as $\frac{FSR(R)}{2FWHM(R)}$ as a function of $R$, (e) WDM channel count $N$ as a function of coupling coefficient $r$.}
    \label{mrr}
\end{figure}

Before proceeding to discussing PhCs, we would like to quantitatively analyse the scaling challenge in MRR-based synapses. The channel count in an MRR-based WDM link is proportional to the ratio of the free-spectral range (FSR) to the full-width half-maximum (FWHM) of the resonance \cite{tait2016microring, tait2016multi}. The design parameters in an MRR available to tune its FSR and FWHM are the radius, $R$ and the gap between the microring and the coupling waveguide which determines the self-coupling coefficient, $r$, (refer Fig. \ref{mrr}(a)). The range of each parameter is constrained by foundry design rules due to their fabrication tolerances. To illustrate the channel count limitation fundamental to MRRs, we show how $R$ and $r$ affect the channel count, $N$ in Figs. \ref{mrr}(b)-(e). Figs. \ref{mrr}(b) and (c) show that FWHM and FSR decrease similarly with increasing $R$. This results in $N = \frac{FSR}{2FWHM}$ being a constant function of $R$, as shown in Fig.\ref{mrr}(d). In other words, one cannot change the allowable channel count by changing the MRR radius. Alternatively, one can vary the self-coupling coefficient, $r$. Fig. \ref{mrr}(e) shows that $N$ increases with $r$. However, one cannot indefinitely increase $r$ --in the high self-coupling limit ($r$ = 0.95), $N$ approaches to 30. This illustrates that MRR-based WDM links run into fundamental limits and cannot be scaled indefinitely.  

% These include: five times higher thermal efficiency, 4 times higher channel count in the C-band, 12 times higher compute density and negligible inter-channel coherent interactions.
 
\section{PhC design}
\begin{figure}[htbp]
    % \centering
    \includegraphics[width=0.5\textwidth]{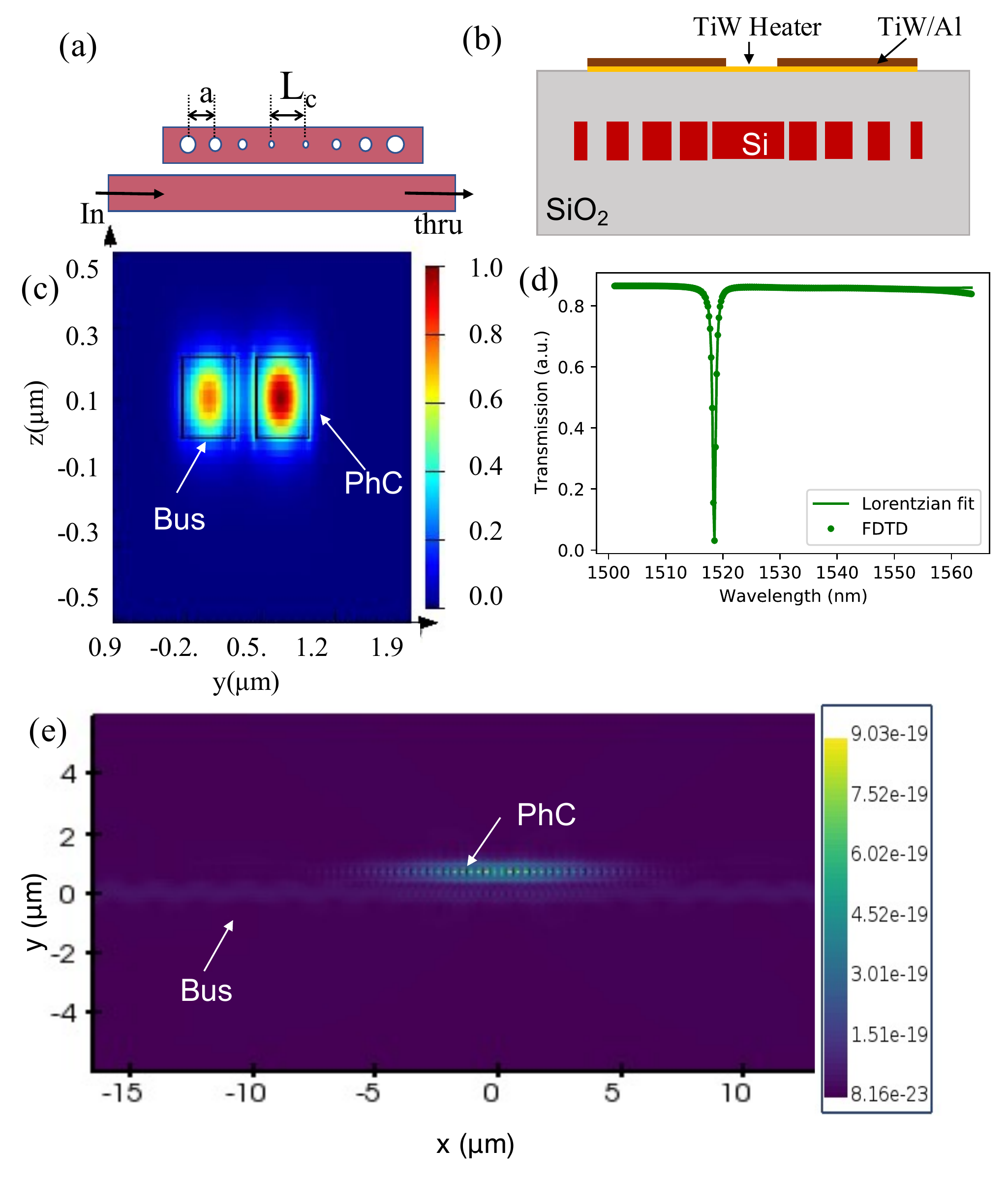}
    \caption{(a) Design schematic of the PhC nanobeam cavity with horizontally symmetric tapered holes with periodicity $a$, and cavity length $L_c$.  Only 4 holes on either sides are shown for simplicity of illustration. (b) Illustration of the PhC cross-section. (c) Transverse mode profile of a bus waveguide and PhC cross-section at the horizontal symmetry plane showing evanescent coupling into the PhC. (d) Simulated transmission spectrum of the PhC cavity, revealing a single Lorentzian resonance feature and thus its FSR-free property. (e) Simulated 2D electric field intensity profile showing the resonant cavity mode confined in the PhC.}
    \label{design}
\end{figure}

An illustrative top-down view of the PhC nanobeam synapse is shown in Fig. \ref{design}(a). It comprises a one-dimensional lattice of cylindrical holes etched in a single-mode silicon waveguide, as evident in the side-view of the cavity in Fig. \ref{design}(b). The 1D nature eases both the design and fabrication constraints relative to higher-dimensional PhC cavities. The PhC is horizontally symmetric with 40 linearly tapered holes with radii in the range [0.11, 0.04] $\mu$m and constant hole spacing of $a = 0.33 \mu$m on either side. The combination of linear radii tapering and a large hole count ensures the phase velocity is preserved within each segment of the nanobeam, minimizing losses in the cavity \cite{deotare2009high, quan2010photonic}. The resonance wavelength of the PhC can be varied by changing the cavity length, $L_c$, which refers to the separation between the two central holes. Light is coupled into the PhC through a bus waveguide as shown in Fig.\ref{design}(a). The gap between the waveguide and the PhC cavity was chosen to be 0.2 $\mu$m, which optimized the cavity quality factor and is close to parameters in \cite{dong2018ultra}. The width and thickness of both the bus and nanobeam waveguides are 500 nm and 220 nm respectively. 
%The effective indices of the Bloch mode $\frac{\lambda}{(2a)}$ match that of the single-mode waveguide, 2.4.  

We simulated the optical field distribution and transmission of the PhC shown in Fig.\ref{design}(a). Fig.\ref{design}(c) shows the optical mode at the horizontal symmetry plane of the PhC, calculated by a finite-difference eigenmode (FDE) solver on Lumerical MODE. It illustrates the evanescent coupling of light from the bus waveguide into the photonic crystal cavity. We then used a variational finite difference time domain solver in Lumerical, to simulate the transmission spectrum of the PhC across a wavelength range of 100 nm. Fig.\ref{design}(d) shows the transmission spectrum of the PhC, showing a single resonance at 1.519$\mu$m. This illustrates the FSR-free nature of the PhC within the wavelength range.  Fig. \ref{design}(e) shows the resonant mode within the cavity, occupying 0.16 $\mu m^3 \approx 0.59 (\lambda/n)^3$, which is at least an order of magnitude lower than typical microrings. Time aphodization was used so that only the resonant mode profile was captured. 
%with calculated quality factor of 3000

For application as synapses, the PhC resonance wavelength needs to be tunable. Similar to MRR-based synapses, tuning the resonance wavelength tunes the transmission value at a given wavelength. In neural network terms, the transmission is the ``weight" associated to a synapse. We employ a simple thermal tuning mechanism using metal heater deposited atop the photonic cavity, as shown in Fig. \ref{design}(b). The heat from the metal heater results in a local temperature increase which tunes the local index, due to the thermo-optic effect, and thereby tunes the cavity resonance. Any other index tuning mechanism can also be employed for this purpose.      

\section{Experimental characterization}
To establish the performance merits of PhCs relative to MRRs, we fabricated devices with the design presented above and experimentally characterized their resonance properties. We experimentally verify that PhCs can be FSR-free within operational wavelength range and also show they have higher tuning efficiency than MRRs. We also discuss the design strategy for designing PhC-based WDM photonic synapses, shown in Figure \ref{fig:model}(c).  

\begin{figure}[htbp]
    \includegraphics[width=0.5\textwidth]{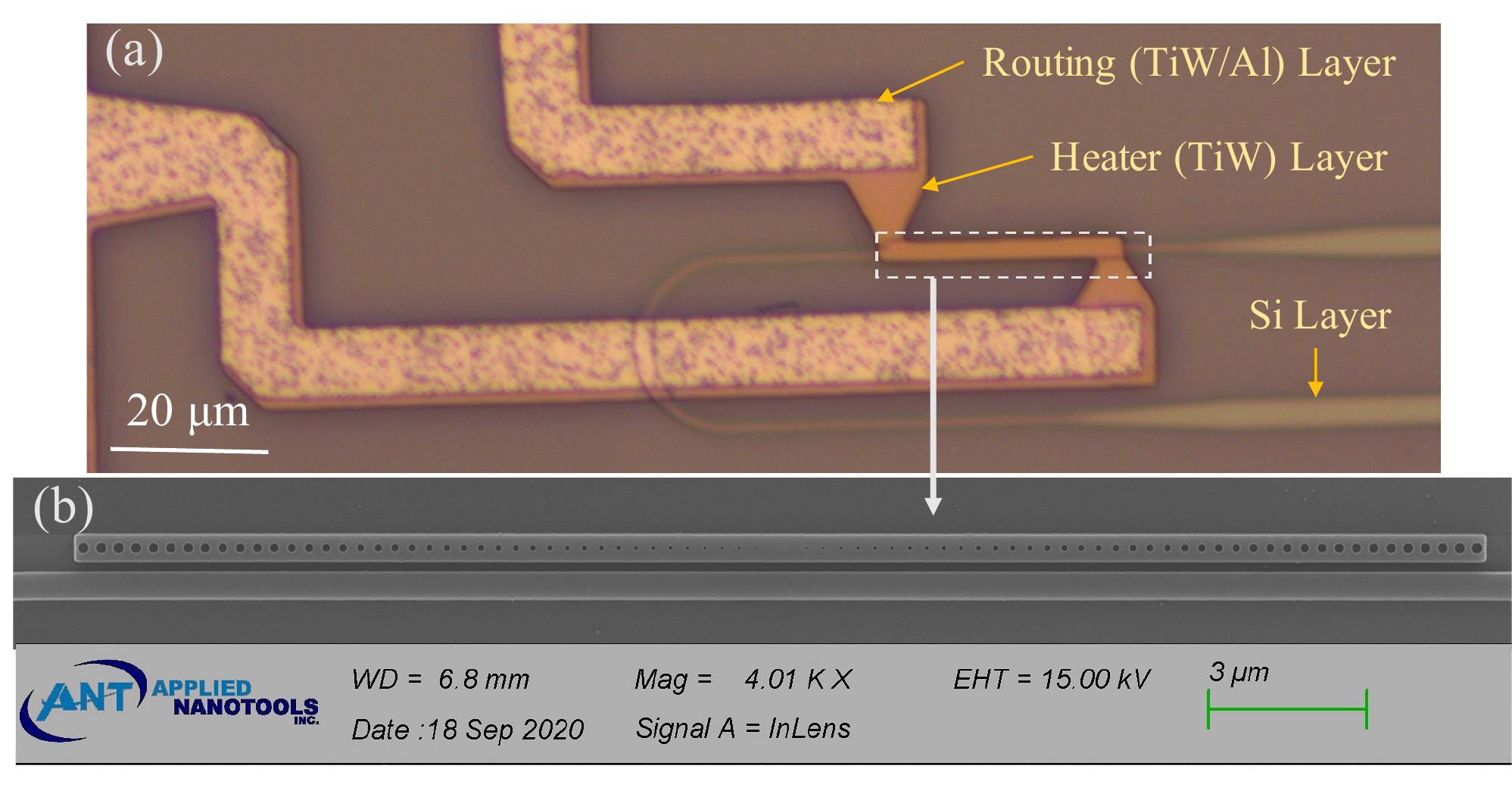}
    \caption{(a) Optical micrograph of a fabricated photonic crystal nanobeam cavity, showing the Si device layer as well as the routing and heater layers (taken with a Zeiss Axioscope microscope) (b) Zoomed-in view of an all-pass photonic crystal nanobeam filter test structure captured with an SEM (Image taken by Applied Nanotools, Canada)}
    \label{fig:sem}
\end{figure}

Fabrication was carried out on a commercial process at Applied Nanotools, Canada. The devices were fabricated on a silicon-on-insulator (SOI) wafer with 3$\mu$m oxide cladding atop silicon device layer. The silicon PhC, shown in the SEM image in Fig.\ref{fig:sem}(b), was patterned via 100 keV electron beam lithography as the feature size entails high resolution fabrication. Fig. \ref{fig:sem}(a) shows the metal layers deposited atop for thermal tuning. A 200 nm thick layer of high-resistance titanium-tungsten alloy (TiW), with bulk resistivity ($\rho$) of 0.61 $\mu \Omega$-m, is used as the heater. A low resistance routing layer composed of tungsten/aluminum bi-layer is used to form contact to the heater. 

\begin{figure}[htbp]
    \centering
    \includegraphics[width=0.5\textwidth]{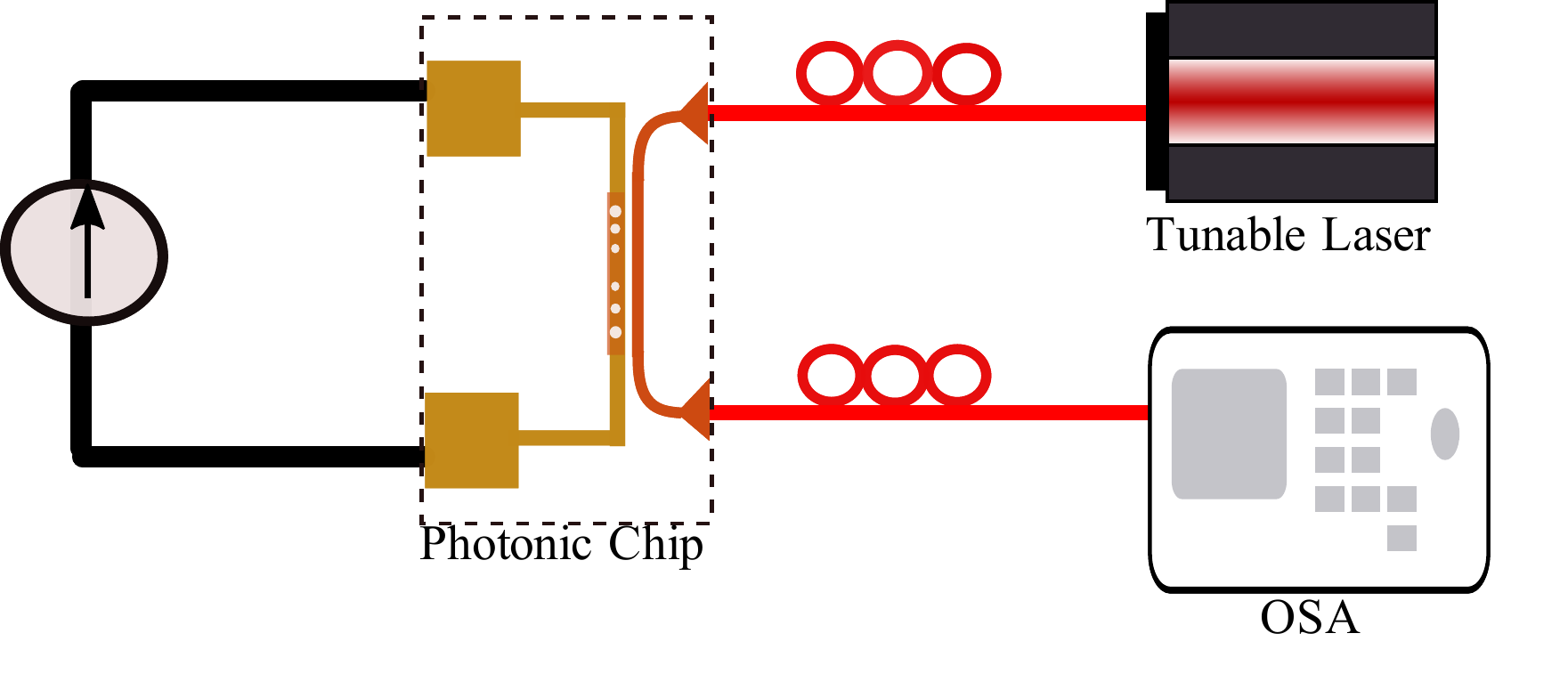}
    \caption{Experimental setup for PhC characterization: electrical control for resonance tuning by a Keithley current source, electrical and optical coupling with the photonic chip is via electrical probes and Vgroove fiber array respectively (not shown). Optical source is from a tunable laser and the output is collected by a optical spectrum analyzer (OSA).}
    \label{fig:setup}
\end{figure}

\indent The schematic of the experimental setup is shown in Fig. \ref{fig:setup}. The photonic chip was placed on a temperature controlled copper mount. The optical coupling into the on-chip waveguide is done by a Vgroove fiber array (not shown) incident on on-chip TE grating couplers. The optical input to the PhC is from a tunable laser (TL) source, and the optical output is fed to an optical spectrum analyzer (OSA).The round-trip optical insertion loss of the grating couplers was 16 dB. For electrical control of the metal heater, an off-chip current source is used, electrical coupling to the on-chip contact pads is through electrical probes (Picoprobe). The current source is remotely controlled using a lab automation software package, Lightlab \cite{lightlab}. 

\begin{figure}[htbp]
    \centering
    \includegraphics[width=0.4\textwidth]{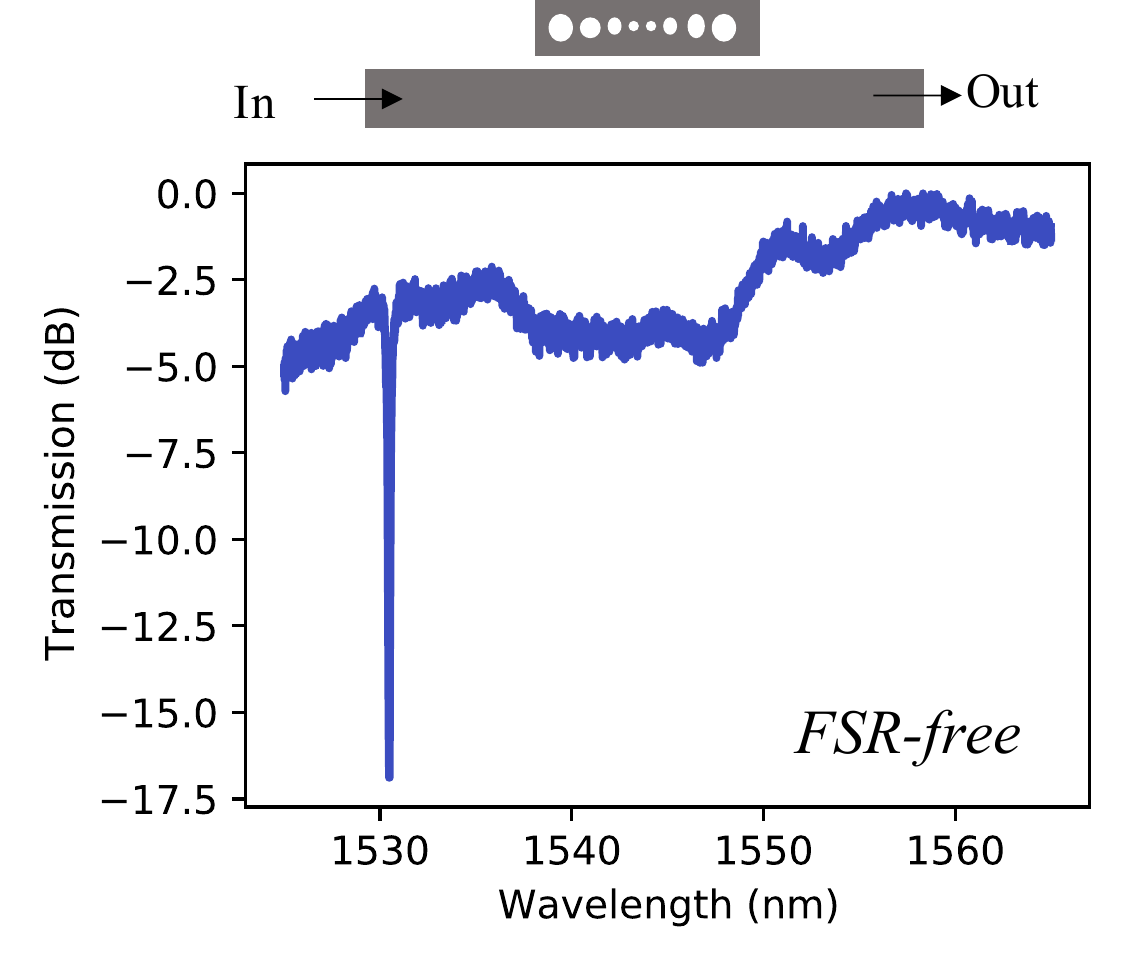}
    \caption{(top) Schematic illustration of the experiment showing the device input and output. (bottom) Measured transmission spectrum of a PhC cavity showing a single resonance within the C-band, verifying its FSR-free property.}
    \label{fig:trans}
\end{figure}

Fig. \ref{fig:trans} shows a schematic illustration of the experiment -- input light of varying wavelengths is sent to the PhC and the output is measured. The measured transmission is thus the ratio of the output to the input at each optical frequency in the range 1525-1565 nm (spanning the entire C-band). From the measured transmission spectrum of a PhC cavity shown, we see a single resonance feature at around 1530.5 nm, with a Q-factor of 12,000 and extinction ratio of $\approx$14 dB. Having only a single resonance indicates the cavity is free-spectral range (FSR)-free. In contrast, for an MRR, a series of recurring resonances spaced an FSR (tens of nm) apart is expected.  \\


\begin{figure}[htbp]
    \centering
    \includegraphics[width=0.5\textwidth]{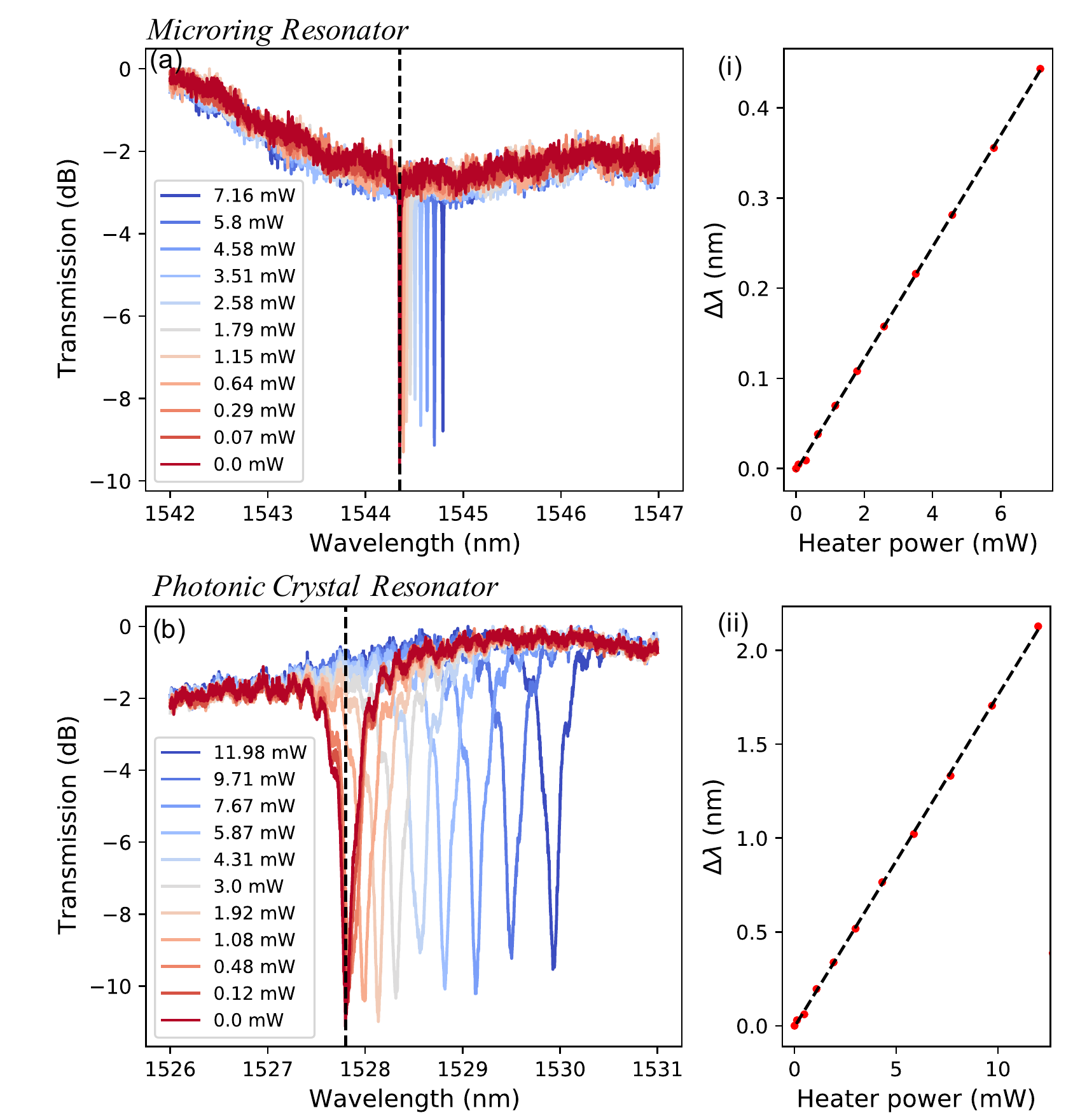}
    \caption{Comparison of thermal tuning efficiency between an MRR and a PhC cavity test structures fabricated on the same chip. (a), (b) show the measured transmission spectra, and (i), (ii) show the corresponding resonance shifts as a function of heater powers for an MRR and a PhC respectively.}
    \label{tuning}
\end{figure}

 We then compare the experimentally measured tuning efficiencies of a PhC and an MRR fabricated on the same photonic die with identical routing and heater metal layers. The MRR design parameters are: 8$\mu$m radius and 300 nm gap between bus and MRR. Tuning efficiency here is defined as the ratio of measured shift of the resonance wavelength ($\Delta \lambda$) to the applied heating power ($P_{\text{heat}}$) calculated as $P_{\text{heat}} = I^2\times R$ where $I$ is the applied current and $R$ is the resistance measured by the Keithley source meter. Figs. \ref{tuning}(a) and (b) show the change in the transmission spectra with thermal tuning for an MRR and a PhC respectively. Fig. \ref{tuning}(i),(ii) show the corresponding resonance shifts as a function of heater power. Using linear regression, we calculate tuning efficiencies of 0.18 nm/mW for the PhC and 0.04 nm/mW for the MRR. The efficiencies are expected to be an underestimate as $R$ is the resistance of the entire circuit. However, it is adequate for a relative comparison, where we find the PhC has a 4.5 times higher thermal tuning efficiency relative to the MRR. Index-tuning efficiency increase here arises from the cavity mode volume reduction in a PhC which translates to an enhanced light-matter interaction relative to an MRR. Ref. \cite{zhou2017compact} demonstrated a three times better tuning efficiency in a PhC nanobeam filter compared to an MRR; our results qualitatively agree. 

%  We also characterized the inter-channel crosstalk in the weight bank shown in Fig.\ref{fig:sem}(a). The PhC filters resonances in the fabricated weight bank were not optimally designed to ensure uniform spacing between the resonances. We investigated the coherent interactions-induced crosstalk between two adjacent PhC filters in the weight bank. 

\subsection{A PhC weight bank}
\begin{figure}
    \centering
    \includegraphics[width =0.5\textwidth]{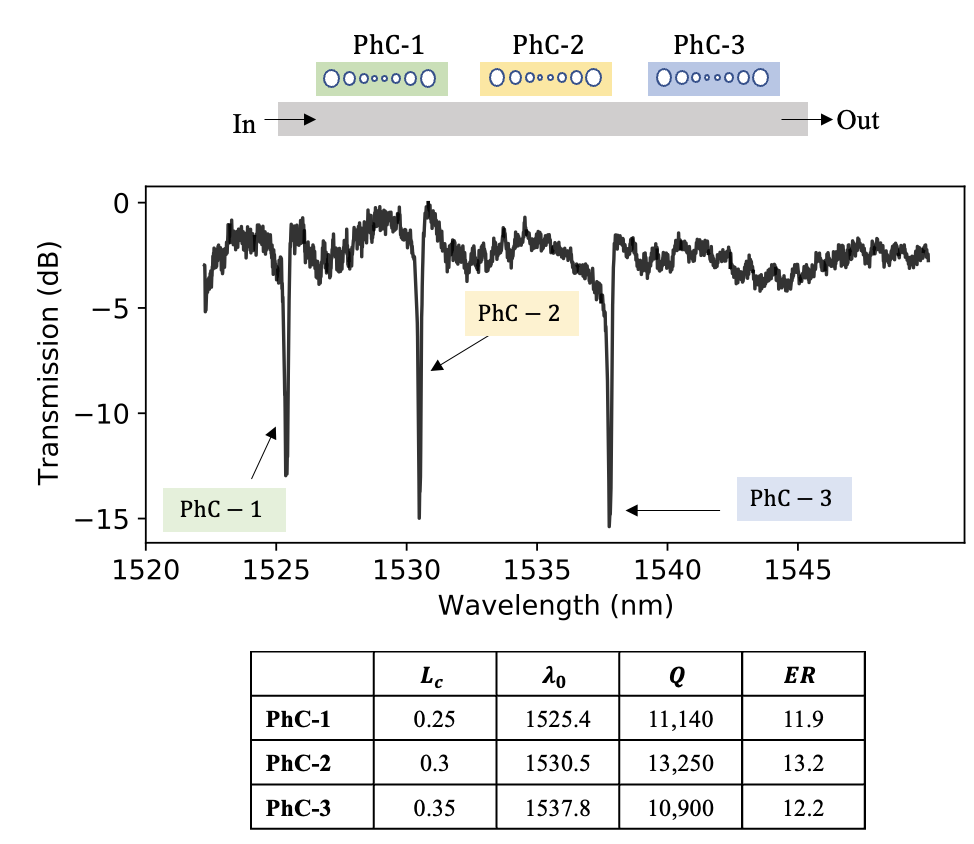}
    \caption{(top) Schematic illustration of a PhC-based WDM weight bank, where each PhC is assigned a different wavelength channel, (middle) Weight bank transmission spectrum, obtained by superposing measured transmission spectra of 3 PhCs, (bottom) Properties of the PhCs, including the cavity length $L_c$, resonance wavelength $\lambda_0$, quality factor $Q$ and extinction ratio $ER$.}
    \label{fig:wb-sim}
\end{figure}

Fig. \ref{fig:wb-sim} illustrates the PhC weight bank design -- it is a series of PhCs coupled to a common bus waveguide. Each PhC of the weight bank is assigned a different wavelength channel, which means the resonances of the PhCs need to varied such that they collectively span the operational wavelength range. One way to design PhCs with varying resonances is to vary the cavity length, $L_c$ (refer Fig. \ref{design}). We fabricated PhCs with varying $L_c$ and measured their transmission spectra and their resonance wavelengths. Fig. \ref{fig:wb-sim} shows the measured transmission spectra of three PhC filters with varying $L_c$ (0.25 $\mu$m, 0.3 $\mu$m and 0.35 $\mu$m). We find that varying $L_c$ by 100 nm results in a resonance wavelength shift of 12.4 nm (as reported in Fig. \ref{fig:wb-sim}). While we didn't have a fabricated weight bank to measure, superposing three measured transmission spectra from three PhCs as shown in Fig.\ref{fig:wb-sim} is a good approximation of the weight bank transmission spectrum. The fabrication limit of the foundry was reported to be about 90 nm, which means varying $L_c$ cannot be a reliable design strategy for finer tuning of the resonance wavelength. For finer tuning, active resonance tuning mechanisms, such as thermal tuning, or post-fabrication trimming \cite{8276286} will be needed.  

\section{Performance benefits of PhC-based synapses}

Here we discuss the system performance metrics where PhC-based synapses outperform their MRR counterparts. Fig. \ref{fig:perf} encapsulates how the fundamental synapse properties lead to overall performance metric improvements. Firstly, as shown in previous sections, PhCs can be designed to be FSR-free, which increases the channel count in WDM-based weight banks. More channels, in turn, means higher information throughput and amenability to high-dimensional inputs. Secondly, PhC nanobeam cavities, as discussed prior, occupy the same area as a single-mode waveguide and have a higher optical confinement. This size reduction brings about benefits in terms of improved tuning energy efficiency and compute density. A quantitative comparison between PhCs and MRRs across these metrics can be found in Table \ref{comptable}.

\begin{table}[]
\centering
\caption{Performance metric comparison between PhCs and MRRs}
\begin{tabular}{|c|c|c|c|}\hline
\multirow{2}{*}{Performance metrics}  &  \multicolumn{2}{c|}{Devices} & Source  \\ 
 & PhC & MRR & \\ \hline
   
Energy per synapse [pJ/bit] &       0.012        &       0.06 & \cite{zhou2017compact}, \cite{7134743}  \\ \hline
Compute density [PMAC/smm$^2$] &      0.77                &   0.1   & Calculated               \\\hline
Throughput (in C-band) [Tbps]   &           7.8         &        1.8     & Calculated      \\\hline
Input dimension (in C-band) & 130 &  30 & Calculated  \\\hline
\end{tabular}
\label{comptable}
\end{table}

\indent First, we examine how eliminating the FSR limitation in PhC increases the WDM channel count relative to MRRs within the C-band. In this comparison, we neglect any inter-channel crosstalk and thus estimate the theoretical channel count in the limit where the channel count, $N = \frac{FSR}{2 FWHM}$ (refer Fig. \ref{mrr}). Due to the FSR-free nature of the PhC filters in the C-band, channels can span the entire C-band. We assume identical Q factor of ~$10^4$, which is typical for silicon MRRs and also within the range of that of our fabricated PhCs as reported in Fig. \ref{fig:wb-sim}. Identical Q implies identical FWHM, which means $N$ now depends solely on the FSR. Assume a typical MRR FSR of 12 nm and we set the FSR proxy of PhC to be 35nm (entire C-band). We find that $N = 130$ for PhCs, which is 4 times higher than in an MRR weight bank (refer Fig. \ref{mrr}). If we don't constrain to the C-band, the system operational wavelength range will only be limited by the bandwidth of other components like off-chip amplifiers, grating couplers, etc. Also, here we assume identical Qs, however by tweaking the mirror holes in the PhC cavity, Q factors up to $10^7$ have been shown \cite{deotare2009high} which would result in $>10^4$ channels just within C-band. We do note that a higher Q means a longer cavity photon lifetime ($\tau$) (by the relation $\tau = \frac{2Q}{\omega_0}$ where $\omega_0$ is the cavity resonance frequency) which can adversely affect other metrics like processing speed. 

% Another practical constraint to scaling up the channel count can be the insertion loss of each PhC -- the loss can be minimized to about 0.4 dB \cite{hendrickson2014ultrasensitive}, which we estimate can accommodate $>$100 channels assuming a detector with 0.8 A/W responsivity.  \\
A direct implication of improved channel count is seen in throughput. Throughput is a primary incentive for using WDM-based photonic synapses. It can be estimated as the signal speed (limited by the synapse bandwidth) times the number of WDM channels. For Q of $10^4$, typical for silicon MRRs, the maximum signal rate would be 60 Gbps. Then the maximum system throughput with the maximum channel count of 30 would be 1.8 Tbps. In PhCs, owing to increased channel count, the throughput can scale upto 7.2 Tbps for C-band.       

For neural networks, the ability to handle high-dimensional input is crucial, especially for applications like image classification (with large pixel count) or natural language processing (NLP). In NLP, a large vector space is required to learn associations between words \cite{li2018word}. Similarly, in image classification applications, the input dimension scales with the image pixel count \cite{bangari2019digital}. Increased channel count using PhCs directly translates to photonic neural networks being amenable to applications with high-dimensional inputs. While dimensionality reduction techniques are often used in neural network models to ease the computational cost, training time and prevent overfitting, the dimensions are still beyond what what MRR-based photonic systems would be able to handle. Without a scalable solution like PhCs, applications with high input dimensions would not be feasible.

\indent 
Energy consumption is often cited as the impetus to novel neural network hardware, in particular photonics \cite{shastri2021photonics}. It is also estimated that the majority of energy consumption in a neural network stems from the energy costs of synapses, since they scale as $n^2$ whereas the nonlinear unit scales as $n$ for $n$ neurons \cite{nahmias2019photonic}. Energy consumption in synapses is the energy associated with tuning and holding the weight value. We experimentally demonstrated that a PhC had about 4.5 times higher thermal tuning efficiency than an MRR (refer Fig. \ref{tuning}). Energy consumption per weighting operation, in units of J/bits, can be calculated as the power consumed divided by the signal speed. With higher tuning efficiency, we can expect proportionately lower energy consumption per weighting operation. The overall energy reduction would then scale with the size of the neural network. The energy consumption values for a single PhC-based synapse and an MRR-based synapse, based on reported thermal tuning efficiency values in prior literature and a bit rate of 10G, are shown in  Table \ref{comptable}.    
Energy efficiency for any index tuning in a photonic cavity depends on the interaction between light and the medium. For photonic cavities, this relationship can be quantified by the Purcell factor, $F_p$, defined as 
\begin{equation}
    F_p  \approx \lambda_0^3 \frac{Q}{V}
\end{equation} where $\lambda_0$ is the cavity resonant wavelength, $Q$ is the cavity quality factor and $V$ is the cavity mode volume. Let $\Delta_p$ be the ratio of the tuning efficiency of a PhC to that of an MRR. We can now calculate $\Delta_p$ as:  
\begin{equation}
    \Delta_p = \frac{F_{p, \text{PhC}}}{F_{p, \text{MRR}}} = \frac{Q_{\text{PhC}}}{Q_{\text{MRR}}}\frac{V_{\text{MRR}}}{V_{\text{PhC}}}. 
    \label{delp}
\end{equation}   
% From Fig. \ref{tuning}, we calculate $Q_{\text{PhC}} = 10^4$, $Q_{\text{MRR}} = 2\times 10^5$ and $\Delta_p = 4.5$, which when plugged into Eq. \ref{delp} gives:   $\frac{V_{\text{MRR}}}{V_{\text{PhC}}} = 90.$ 
If we can assume $Q_{\text{PhC}} = Q_{\text{MRR}}$, the tuning efficiency of PhC is proportionately reduced with the cavity mode volume relative to an MRR. Since we had $Q_{\text{PhC}} =\frac{1}{20} Q_{\text{MRR}}$ in our experiment, the tuning efficiency improvement was not too significant. PhC cavities with $Q \approx 10^7$ have been demonstrated before \cite{deotare2009high}, meaning PhCs should offer orders of magnitude higher tuning efficiency relative to MRRs.  
Another point to note is that this energy efficiency improvement translates to any index tuning mechanism, for example carrier modulation \cite{xu200712}. 

\indent Finally, we compare MRRs and PhCs against another important metric, compute density. It is defined as the numbers of operations done per unit time per unit MAC area, where MAC refers to the multiply-and-accumulate operation in a synapse unit \cite{nahmias2019photonic}. In simple terms, it is a measure how physically compact can a processor be. It is also useful as a standard metric to compare different types of neural hardware (photonic, digital and analog electronics), which vary greatly in size and multiplexing techniques. In \cite{nahmias2019photonic}, the authors estimated that sub-$\lambda$ photonics can achieve 100 times higher compute density than MRR-based architectures. For a typical MRR of radius 7 $\mu$m \cite{tait2016multi} and a sampling speed of 10 GS/s, its compute density would be 100 tera MAC/smm$^2$. For the PhC filter presented here, with footprint of 13 $\mu$m$^2$ and the same sampling speed, the compute density is 770 tera MAC/smm$^2$, about 8 times higher than an MRR. The limit to the physical compactness would eventually come from thermal crosstalk between neighboring devices. However, clever design can alleviate this issue. In Fig. \ref{design}(e), we see that the optical mode is actually confined to a much smaller area than the PhC length. This means that the heater trace need only be as large as the optical mode, which would increase the physical separation between heaters of neighboring PhCs and minimize any inter-channel thermal crosstalk.  Compute density is critical for enabling large-scale photonic neural networks on edge devices where footprint can be a limiting factor. 

While our results are promising, further investigation will be necessary to enable photonic crystals as synapses in photonic neural networks. For example, it is worthwhile to look into ways to increase the PhC quality factor which will improve the tuning energy efficiency.  Next, we can test the synapse architecture in Fig. \ref{fig:model}(c), specifically, demonstrate multi-wavelength weighting by PhCs, summing by a photodetector and a nonlinear transformation by an electro-optic modulator. Also, the current PhC synapse architecture, as shown in Fig. \ref{fig:model}(c), only allows for positive weights, unlike the add-drop MRR synapse architecture which permits both positive and negative weights. Positive weights suffice for typical artificial neural networks, but spiking networks require negative weights for inhibition  \cite{prucnal2017neuromorphic}. Future work can explore improved PhC synapse architecture that can allow for negative weights in spiking networks. Finally, the tuning speed of thermal heaters is limited to 100 kHz \cite{6336779}, for higher tuning speeds, carrier-based index modulation using PhCs will need to be explored.

\section{Conclusion}
\indent High bandwidth of photonics is what makes it incredibly powerful, both for signal communication and processing. To take advantage of the bandwidth, photonic systems often implement wavelength division multiplexing (WDM). Prior work on WDM-based photonic neural hardware uses microring resonators as the synapses, which fundamentally have an upper limit on the channel count, that prohibits network scale. We propose a photonic crystal nanobeam cavity based synapse  as the most scalable synapse solution for photonic neural networks. We have experimentally demonstrated superior performance merits of PhCs pertinent to synaptic weighting including: first, an FSR-free operation, which results in 4 times higher channel density within the C-band and second, a 4.5 times higher weight tuning efficiency resulting from its smaller mode volume. Increased channel count improves the system throughput as well as makes the hardware amenable to applications with high-dimensional input like natural language processing. Another benefit from the smaller footprint of PhC is in compute density, a measure of processing done per unit physical area, which we estimate to be about 8 times higher compute density for our current design. Further work will be dedicated to cavity design optimization, and development of calibration and control procedures for practical applications of such weight banks.    

\bibliographystyle{IEEEtran}
\bibliography{manuscript}

\end{document}